# Magnetic Bubble Domain Blowing with Electric Probe

D.P. Kulikova[1], A.P. Pyatakov [1,2,a], E.P. Nikolaeva[1], A.S. Sergeev[1], T.B. Kosykh[1],

Z.A. Pyatakova[1], A.V. Nikolaev[1], A.K. Zvezdin[2,3]

[1)] Physics Department, M.V. Lomonosov MSU, Leninskie gori, Moscow, 119991, Russia

[2)] A. M. Prokhorov General Physics Institute, 38, Vavilova st., Moscow, 119991, Russia

3) Moscow Institute of Physics and Technology (State University) - Institutskii per. 9, Dolgoprudny, 141700, Russia

[a]pyatakov@physics.msu.ru;

**Abstract.**

The problem of bubble domain generation is revisited from a new perspective that was opened by recent findings in the field of physics of magnetic skyrmions. The single bubble domain can be generated under the tip electrode touching the surface of dielectric iron garnet film by positive step-like voltage pulse and its equilibrium size can be increased with further rising of electric potential. The theoretical analysis shows that the same mechanism can be used to stabilize the submicron bubbles topologically equivalent to the skyrmion.

## Introduction

The emergent field of magnetic skyrmion physics [1]–[3] makes a new look at a time-honored physics of magnetic bubble domains. In the conventional theory of bubble domains there are four interactions that govern the stability of bubble domain state: magnetic exchange interaction, magnetic anisotropy, dipole-dipole interaction, and Zeeman energy in external field. Meanwhile for magnetic skyrmion state the central role in the competition with exchange energy is played by Dzyaloshinskii-Moriya interaction [2]. Recent experiments in iron garnet films [4]–[6] show that the so-called flexomagnetoelectric effect (electric polarization generated by magnetization distribution bending [7], [8]) that also stems from the Dzyaloshinskii-Moria interaction [9] plays important role in micromagnetism.

Worth noting that skyrmions in chiral magnets are represented not only in the skyrmion lattice phase, but also as individual objects [10]. The controlled generation of single skyrmions has been demonstrated recently using scanning tunneling microscopy technique [11][12]. It was shown that in ultrathin films the tip probe can locally create the conditions for nucleation of single skyrmion both by spin polarized current [11] and electric field [12]. Several mechanisms of the electric field induced skyrmion nucleation including flexomagnetoelectric interaction [8] and d-p hybridization [13] were proposed.

In this paper it is shown that the electric tip electrode can generate in thick insulating iron garnet films large magnetic inhomogeneity such as bubble domain. The equilibrium size of bubble domain can be reversibly controlled by static electric field that shows the importance of magnetoelectric interaction in the energy balance that governs the stability of single bubble domain. This observation is also of interest in the context of recent theoretical prediction by Dzyaloshinskii [14] that the electric field of the tip can create the domain wall in the single domain magnetic state.

**Experiment**

In our experiment we used the sample with the pronounced magnetoelectric effect [15]. It was the 9.7 μm thick iron garnet film $(BiLu)_3(FeGa)_5(O)_{12}$ grown by liquid-phase epitaxy on (210) $Gd_3Ga_5O_{12}$ substrate. The substrate thickness was 0.5 mm. The period of the stripe domain structure was 29 μm and *$4\pi M_S$*=62 Gs.

The experimental scheme is shown in Figure 1. To produce a high-strength electric field in the dielectric iron garnet film, we used a 10 μm-diameter molybdenum wire with a pointed tip, which touched the surface of the sample. This allowed us to obtain electric field strength up to 1000 kV/cm near the tip by supplying a voltage of 500 V to the needle. High electrostatic field did not cause dielectric breakdown because it decreased rapidly with the distance from the probe and near the bottom of the substrate the strength of the field did not exceed 100 V/cm.

To prevent the leakage currents and suppress the current of charging/discharging we used the ballast resistor of 10 MOhm series-connected to the tip electrode (fig.1). The ballast resistor suppresses the currents to values well below 0.1 mA (that is not enough to induce the changes in our samples) at the same time leaving the voltage jump sharp enough to generate bubble domain (the charging time ~ 100 ms).

The magneto-optical technique in Faraday geometry was used to observe the micromagnetic structure. The image of the magnetic structure was taken by CCD camera.

The magnetic bias field makes the magnetic state bistable: the magnetic domains are about to arise in the homogeneously magnetized magnetic film. The in-plane [$\bar{1}20$] and out of-plane [210] components of the bias field were 200 Oe and 70 Oe, respectively.

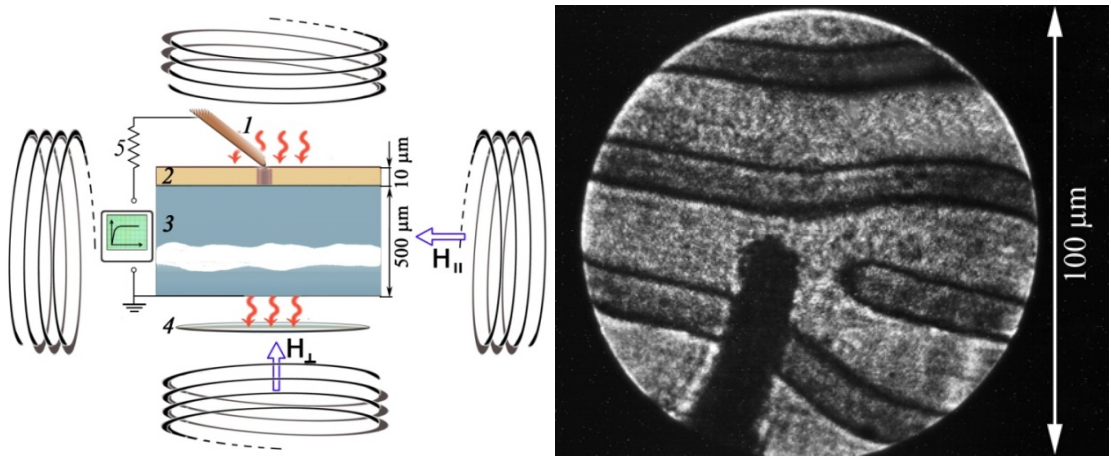

Fig.1. The schematics of the experimental setup: a) 1 is the tip electrode, 2 is the iron garnet film, 3 is the substrate, 4 is the objective lens of microscope, 5 is the ballast resistor. Two pairs of coils create two component bias field: in-plane $H_\parallel$ and out of plane $H_\perp$; b) The view of the stripe domain structure in microscope (bias field is below the collapse value). The black object is the wire of electrode touching the surface.

We observed bubble domains blowing under the electrically charged tip electrode touching the surface of an iron garnet film (fig.2a). When the positive voltage more than 500 V was applied to the tip (actually, the threshold voltage depended on magnetic bias) the bubble domain appeared (fig 2b). No bubble creation was observed at the negative electric bias of the tip electrode. After the voltage was switched off the domain detached from the tip and reduced in size (fig 2c). If the positive voltage is switched again the bubble restores the size and position of fig2 (b). Another bubble can be generated when the tip is moved far from the bubble at the distance of several bubble diameters.

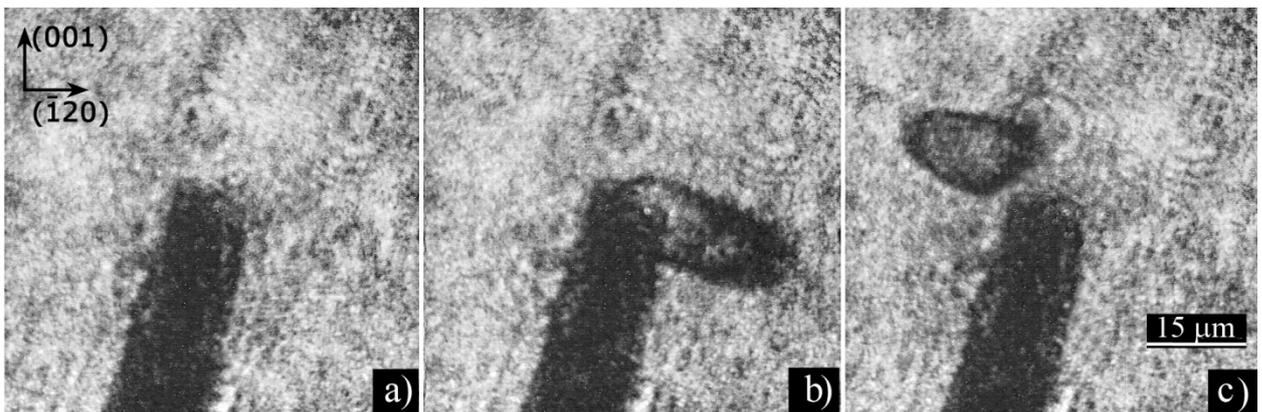

Fig.2. The magneto-optical images of the process of bubble domain blowing with the tip electrode: (a) the tip electrode touching the surface of iron garnet film in single domain state (bias field is higher than the collapse value), there is no voltage at the tip (b) the voltage is switched on, the bubble domain of elliptical shape is seen at the right side of the tip; (c) the bubble domain detaches from the tip when the voltage is switched off.

The size of the newly born bubble enlarges with further increasing of electric voltage (fig 3). It has elliptical shape elongated along the [1̄20] direction and the major semiaxis enlarges with electric voltage until the bubble evolves to stripe domain through elliptical instability (fig 3). The process of bubble inflation was reversible at electric field increasing/decreasing below the critical field of elliptical instability.

The increase of the resistance value of the balance resistor by two orders of value does not change the picture of bubble inflating in the electric field. Since the resistor 500 MOhm dramatically increase the charging time, the jump of the electric field during the switching process is not such a sharp as for 10 MOhm thus preventing the bubble generation process.

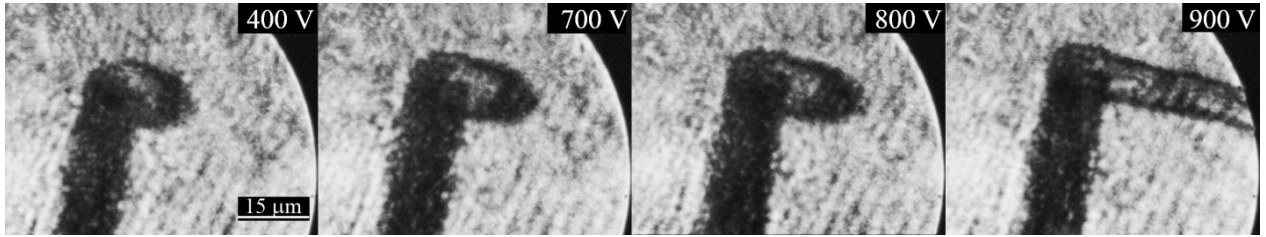

Fig.3. The inflation of the bubble domain and its transformation to the stripe domain with the growth of electric voltage on the tip electrode.

## Discussion

The dependence of the bubble generation on the electric polarity of the tip allows us to rule out the effect of mechanical pressure of the tip since electrostatic attraction to neutral dielectric surface does not depend on the electric polarity. Since we use a paramagnetic tip, the spin-transfer torque mechanism is also eliminated. The leakage current and magnetic field related to it can be ruled out since the variation of balance resistor in the range of two orders does not affect the size of the bubble in the experiments of figure 3.

Some questions remain about the role of the offset current (the current of charging and discharging of the tip). It is known that to generate the bubble domain lattice the bias magnetic field should be switched on abruptly to shake up the structure. Our experiment shows that the same is true for the electric field. The offset current during the charging process is not zero in this case (the rise time is of the order of 100 ms). Thus we cannot eliminate completely the influence of magnetic field generated by the tiny offset current in the nano-sized region of the tip-surface contact. However when the size of bubble is higher than 1 μm this magnetic field is negligible.

It should be emphasized that the necessary condition for bubble domain generation and inflating with the growth of electric field is the in-plane magnetic field. This fact should be related with the previous observation that external magnetic field dramatically enhances the effect of electric-field-induced domain wall motion [16][6]. It can be explained in the framework of concept of flexomagnetoelectricity [8]: the electric polarization appears only in Neel-type domain wall and its sign depends on the sense of magnetization rotation in magnetic domain walls (domain wall chirality). The external in-plane magnetic field reorients the magnetization direction in the center of the domain wall thus making it Neel-like and imposing the certain chirality. In case of the in-plane magnetic field directed along $[\bar{1}20]$ crystal axis (fig 4) the left and right edges of the bubble domain have opposite rotational sense and thus opposite surface electric charges. One edge of the bubble domain is attracted to the tip the other is repelled thus constituting the pair of forces stretching the bubble domain along the $[\bar{1}20]$ crystal axis.

Of special interest are the first moments of the bubble domain generation. The nanometric spin whirls nucleated under the tip is not a skyrmion in the popular sense since it has a zero topological number and is similar to the submicron bubbles with two Bloch lines observed in [17]. However a topologically protected bubble without Bloch lines being 100 nm in diameter is equivalent to skyrmion [17]. According to [18] a skyrmion is an intermediate state of a bubble domain blowing process. Originally "the skyrmion model" [18] was proposed for dynamical bubble domain nucleation in the homogeneous defect-free area in front of advancing ferromagnetic [19] or ferroelectric [20] domain wall. Here we report on the static mechanism of bubble nucleation by electric field from single domain state. Since the nucleation process implies that electric field exerts a torque on magnetic moments it can be speculated that similar mechanism is involved in skyrmion electric field-induced nucleation process observed in ultrathin films of metals [12]. In our case we deal with the isolators so the leakage current driven mechanisms are even more unlikely than in the case of ref. 12.

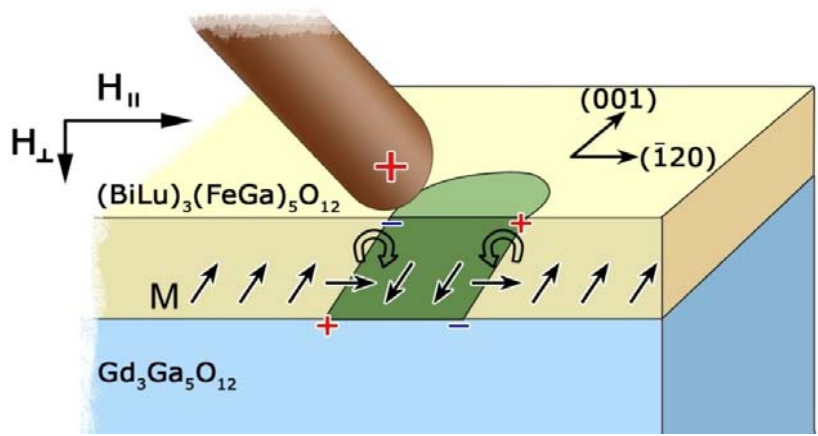

Fig. 4. The mechanism of the effect of bubble domain inflation in electric field observed in figure 3. The black straight arrow shows the magnetization distribution in iron garnet film. The hollow circle arrows show the chirality, i.e. the sense of magnetization rotation if one travels along $(\bar{1}20)$ direction.

In this context let us consider a problem of stabilization of bubble with nonzero topological charge in electric field. This bubble has the Neel-type domain walls without Bloch lines and due to the flexomagnetoelectric effect interact with the electrically charged tip (fig.5).

In classical theory of bubble domains [21] the stability of the bubble with diameter $d$ in magnetic film with the thickness $h$ and magnetization $M_s$ can be described by the balance of the effective magnetic fields:

- the effective field of the domain wall surface tension $\sigma$:  $-\sigma/(M_s d)$

- the magnetic field of magnetostatic interaction expressed in terms of Thiele's force function:

$$+4\pi M_s \frac{h}{d} F\left(\frac{d}{h}\right)$$

- and external magnetic bias field $-H_b$

Here signes "+" and "-" shows the tendency of bubble to expand or shrink, respectively. Due to the presence of surface tension there is a minimum diameter $d_c$ of stable skyrmion. The increase of bias field $H_b$ leads to the collapse of bubble domains.

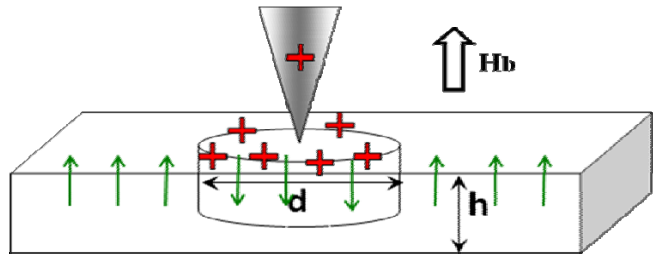

Fig. 5. The problem of submicron bubble (skyrmion) stabilization in the electric field. Green arrows show magnetization direction. $H_b$ is magnetic bias field.

In the presence of external electric field additional flexomagnetoelectric term appears described by effective field:  $H_{me} = \frac{2\gamma}{M_s}\left(\vec{E}\cdot div\vec{m} - grad\left(\vec{E}\cdot\overset{\downarrow}{\vec{m}}\right)\right)$

where the $\gamma$ is flexomagnetoelectric constant, $\mathbf{m}=\mathbf{M}/M_s$ is normalized magnetization vector, the vertical arrow shows that differential operator acts on the magnetization only.

Taking into account that the dependence of electric field on the distance is proportional to $1/r^2$ we can see that this contribution is predominant when $d$ tends to zero. Taking the value $\gamma=10^{-6}$ CGS unit [7], $M_s$=5G, domain wall width ~100nm, film thickness h=10μm, the bias field $H_b$ in the range 10 Oe and the voltage at the tip ~300V, we obtain the characteristic size of stabilized skyrmion in submicron range d<1μm when the tip curvature radius is 30nm and below.

**Conclusion**

Thus it is experimentally shown that electric probe can generate the single bubble domain in a controlled way: at the arbitrary cite of the surface and with the size adjusted by electric field.

The influence of conduction current in the experiment (through the magnetic field or spin torque related to it) was shown to be negligible since the stability of bubble domain and its size does not depend on the value of the balance resistor. The role of the offset current is not clear since the electrical shock (the jump of voltage) necessary for bubble domain generation implies the offset current with nonzero magnetic field. Nevertheless we can positively state that static electric field is responsible for the force that stretches the bubble domain thus facilitating the bubble generation and inflating it to the stable state with the diameter larger than the critical one. This work also provides the experimental evidence for manipulation of bubbles using an electric field that was proposed earlier in [22][23].

The flexomagnetoelectric mechanism described in this paper can be used to generate skyrmions by electric field rather than by current pulses of large density as was reported previously in [11] for STM–probe writing/deleting, or in recent experiments on "blowing magnetic skyrmion bubbles" by geometrical constriction [24]. It also can be the origin of the effect of reversible switching between a single skyrmion and the ferromagnetic state by local electric field from the STM tip recently observed in ultrathin films of magnetic metals [12]. The theoretical estimates show the possibility of the stabilization submicron bubble domain topologically equivalent to the single skyrmion.